\documentclass[twocolumn,showpacs,showkeys,preprintnumbers]{revtex4-1}
\usepackage{amssymb}

\usepackage{amsfonts}

\usepackage{latexsym}

\usepackage{dcolumn}

\usepackage{amsmath}

\usepackage{graphicx}

\usepackage{psfrag,pstricks}

\begin{document}

\title{Dispersive interaction of a Bose-Einstein condensate with the movable mirror of an optomechanical cavity in the presence of the laser phase noise}

\author{A. Dalafi$^{1}$ }
\email{adalafi@yahoo.co.uk}

\author{M. H. Naderi$^{2}$}

\affiliation{$^{1}$ Laser and Plasma Research Institute, Shahid Beheshti University, Tehran 1983969411, Iran\\
$^{2}$Quantum Optics Group, Department of Physics, Faculty of Science, University of Isfahan, Hezar Jerib, 81746-73441, Isfahan, Iran}

\date{\today}

\begin{abstract}

We theoretically investigate the dispersive interaction of a Bose-Einstein condensate (BEC) trapped inside an optomechanical cavity with a moving end mirror in the presence of the laser phase noise (LPN) as well as the atomic collisions. We assume that the effective frequency of the optical mode is much greater than those of the mechanical and the Bogoliubov modes of the movable mirror and the BEC . In the adiabatic approximation where the damping rate of the cavity is faster than those of the other modes, the system behaves as an effective two-mode model in which the atomic and mechanical modes are coupled to each other through the mediation of the optical field by an effective coupling parameter.  We show that in the effective two-mode model, the LPN appears as a classical stochastic pump term which drives the amplitude quadratures of the mechanical and the Bogoliubov modes. It is also shown that a strong stationary mirror-atom entanglement can be established just in the dispersive and Doppler regimes where the two modes come into resonance with each other and the effect of the LPN gets very low. 

\end{abstract}


\maketitle

\section{Introduction}
%
%
Today, thanks to the optomechanical cavities equipped with ultracold atoms, a suitable basement have been provided to study atom-photon interaction in the regime where their quantum mechanical properties are manifested in the same level \cite{Maschler2008, dom JOSA}. Besides, the hybrid systems consisting of Bose-Einstein condensates (BEC) trapped inside the optical lattice of a cavity exhibit optomechanical properties \cite{Bha 2009,Bha 2010,Gupta,Brenn Nature,Kanamoto 2010}. It has been shown \cite{Masch Ritch 2004,Dom JB} that if the laser pump which drives the cavity containing a BEC is far detuned from the atomic resonance, the excited electronic state of the atoms can be adiabatically eliminated and consequently the only degrees of freedom of atoms will be their mechanical motions.

In a situation where only the lowest modes of the matter field are excited (the condensate mode is considered as a classical mean field and the first excited quantum mode is called the Bogoliubov mode), the BEC behaves just like a single-mode quantum oscillator which is coupled to the radiation pressure of the cavity \cite{Nagy Ritsch 2009,dalafi1}. The Bogoliubov mode of the BEC acts effectively as a secondary mechanical mode in the cavities having a moving end mirror \cite{Genes2008,Asjad,dalafi2} or behaves as one of the several membranes inside a cavity which interact with the optical field \cite{Seok}.

The other interesting feature of such hybrid systems containing the BEC, is the nonlinear effect of the atom-atom interaction which can affect optomechanical properties of the system \cite{Zhang 2009}. It may lead to the squeezing of the Bogoliubov mode \cite{dalafi3} and change the pattern of normal modes of the cavity output optical field \cite{dalafi5}. Besides, it can change the threshold of the quantum phase transition of the BEC \cite{dalafi4}. On the other hand, for an interacting BEC the frequency of the Bogoliubov mode is a function of the \textit{s}-wave scattering frequency ($ \omega_{sw} $) of the nonlinear atom-atom interaction \cite{nagy2013}. Since $ \omega_{sw} $ can be manipulated by the transverse trapping frequency of the BEC \cite{Morsch},  the frequency of the Bogoliubov mode is a controllable parameter. It is one of the advantages of the hybrid system with respect to the bare one in which the frequency of the mechanical mode of the moving mirror is a fixed parameter.

In this paper we study the interaction of a BEC trapped inside an optomechanical cavity with a moving end mirror in the dispersive regime where the effective frequency of the optical mode is much greater than those of the mechanical and the Bogoliubov modes. Besides, we consider the situation where the damping rate of the cavity is faster than those of the other modes so that the optical field follows the dynamics of the mechanical and atomic oscillators adiabatically. In this way, the three-mode system behaves as an effective two-mode one so that the two modes interact with each other through the mediation of the optical field. 

Here, we also consider the effect of the atom-atom interaction as well as the classical noise of the external laser which pumps the cavity. The laser phase noise (LPN) have been known as one of the possible technical limitations in realized setups which may affect the cooling and the coherent dynamics in optomechanical systems \cite{Kennedy,Diosi,Rabl2011,Yin,Phelps,Abdi2011,Ghobadi}. We derive an effective Hamiltonian in the adiabatic approximation and show explicitly that the LPN acts as a classical stochastic driving term which pumps the amplitude quadratures of both the mechanical and the Bogoliubov modes.  We also show how the quantum noise of the optical field affects the Brownian noises of these modes. This is one of the main results of the present work. Furthermore, we show that at very low temperature where the environmental thermal noises are negligible, the destructive effect of LPN on the stationary bipartite entanglements of the system is more considerable, especially when the linewidth of the pump laser is not low enough.

On the other hand, due to the indirect interaction of the atomic and the mechanical modes, establishing a strong stationary entanglement between them is very challenging \cite{Rogers}. Nevertheless, production of the atom-field and mirror-field entanglement is straightforward since the atomic and the mechanical modes interact directly with the optical field \cite{Chiara, dalafi2}. As has been shown in Ref.\cite{Rogers}, the atom-mirror entanglement is nonzero only within a very short time window and goes to zero in the steady state when the cavity is pumped by an unmodulated laser. However, we show that a strong stationary mirror-atom entanglement can be established by a cw pump laser if the system works in the dispersive regime where the resonance condition between the mechanical and the Bogoliubov modes is fulfilled. 

The paper is structured as follows. In Sec. II we derive the Hamiltonian of the system consisting of a BEC inside an optomechanical cavity with a moving end mirror considering the \textit{s}-wave scattering interaction and the effect of the laser phase noise. In Sec. III we describe the classical model of the phase noise and in Sec. IV the quantum Langevin equations (QLEs) are derived and linearized around the semiclassical steady state. In Sec. V we derive the effective two mode model of the system by adiabaticaly eliminating the optical field. In Sec. VI the effects of the LPN and the atom-atom interaction on the stationary bipartite entanglements are studied. Finally, our conclusions are summarized in Sec. VII.
\section{System Hamiltonian}

\begin{figure}[ht]
\centering
\includegraphics[width=3in]{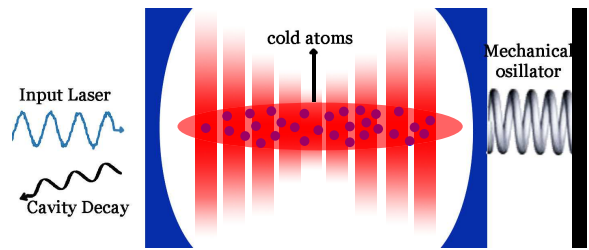} 
\caption{(Color online) A BEC trapped in an optomechanical cavity interacting dispersively with a single cavity mode. The cavity which decays at rate $\kappa$ is driven through the fixed mirror by a laser with frequency $\omega_{p}$ whose phase fluctuates randomly and the end mirror is free to oscillate at mechanical frequency $\omega_{m}$.}
\label{fig:fig1}
\end{figure}

We consider an optomechanical cavity with length $L$ whose end mirror is free to oscillate at mechanical frequency $\omega_{m}$. The cavity consists of a cigar shaped BEC of $N$ two-level atoms with mass $m_{a}$ and transition frequency $\omega_{a}$ confined in a cylindrically symmetric trap with a transverse trapping frequency $\omega_{\mathrm{\perp}}$ and negligible longitudinal confinement along the $x$ direction (Fig.\ref{fig:fig1}).

The cavity is driven at rate $\eta=\sqrt{2\mathcal{P}\kappa/\hbar\omega_{c}}$ through the fixed mirror by a laser with frequency $\omega_{p}$, and wavenumber $k=\omega_{p}/c$ ($\mathcal{P}$ is the laser power and $\kappa$ is the cavity decay rate). In the dispersive regime where the laser pump is far detuned from the atomic resonance ($\Delta_{a}=\omega_{p}-\omega_{a}$  exceeds the atomic linewidth $\gamma$ by orders of magnitude), the excited electronic state of the atoms can be adiabatically eliminated and spontaneous emission can be neglected \cite{Masch Ritch 2004,Dom JB,Nagy Ritsch 2009}. In this way, we can describe the dynamics within an effective one-dimensional model by quantizing the atomic motional degree of freedom along the $x$ axis only. The Hamiltonian of the system can be written as
\begin{eqnarray}\label{H1}
H&=&\hbar\omega_{0} a^{\dagger} a + i\hbar\eta[a e^{i\omega_{p}t}e^{i\phi(t)}-a^{\dagger} e^{-i\omega_{p}t}e^{-i\phi(t)}]\nonumber\\
&&+\int_{-L/2}^{L/2} dx \Psi^{\dagger}(x)\Big[\frac{-\hbar^{2}}{2m_{a}}\frac{d^{2}}{dx^{2}}+\hbar U_{0} \cos^2(kx) a^{\dagger} a\nonumber\\
&&+\frac{1}{2} U_{s}\Psi^{\dagger}(x)\Psi(x)\Big] \Psi(x)+\frac{1}{2}\hbar\omega_{m}(p^{2}+q^{2})-\hbar\xi_{m} a^{\dagger}a q.\nonumber\\
\end{eqnarray}

Here, $ \omega_{0} $ and $ a $ are the frequency and the annihilation operator of the cavity mode. The second term describes the cavity pumping term with a laser whose phase variation in time is given by $ \phi(t) $ which is considered as a classical stochastic variable. $U_{0}=g_{0}^{2}/\Delta_{a}$ is the optical lattice barrier height per photon which represents the atomic backaction on the field, $g_{0}$ is the vacuum Rabi frequency, $U_{s}=\frac{4\pi\hbar^{2} a_{s}}{m_{a}}$ and $a_{s}$ is the two-body \textit{s}-wave scattering length \cite{Masch Ritch 2004,Dom JB}. The last two terms in the Hamiltonian of Eq.(\ref{H1}) represent, respectively, the energy of the mechanical mode and the radiation pressure coupling of rate $\xi_{m}=(\omega_{0}/L)\sqrt{\hbar/m\omega_{m}}$ where $m$ is the effective mass of the moving mirror.

In the weakly interacting regime,where $U_{0}\langle a^{\dagger}a\rangle\leq 10\omega_{R}$ ($\omega_{R}=\frac{\hslash k^{2}}{2m_{a}}$ is the recoil frequency of the condensate atoms), and under the Bogoliubov approximation \cite{Nagy Ritsch 2009}, the atomic field operator can be expanded as the following single-mode quantum field
\begin{equation}\label{opaf}
\Psi(x)=\sqrt{\frac{N}{L}}+\sqrt{\frac{2}{L}}\cos(2kx) c,
\end{equation}
where the so-called Bogoliubov mode $ c $ corresponds to the quantum fluctuations of the atomic field about the classical condensate mode ($ \sqrt{\frac{N}{L}} $). By substituting the atomic field operator of Eq.(\ref{opaf}) into Eq.(\ref{H1}), the Hamiltonian of the systems reduces to
\begin{subequations}
\begin{eqnarray}
H&=&\hbar\tilde{\omega}_{0} a^{\dagger} a + i\hbar\eta[a e^{i\omega_{p}t}e^{i\phi(t)}-a^{\dagger}e^{-i\omega_{p}t}e^{-i\phi(t)}]\label{subH}\nonumber\\
&&+\frac{1}{2}\hbar\omega_{m}(p^{2}+q^{2})-\hbar\xi_{m} a^{\dagger}a q + H_{c},\\
H_{c}&=&\hbar\Omega_{c} c^{\dagger}c+\frac{1}{4}\hbar\omega_{sw}(c^{2}+c^{\dagger 2})+\frac{\hbar}{\sqrt{2}}\zeta a^{\dagger}a (c+c^{\dagger}),\nonumber\\
\end{eqnarray}
\end{subequations}
where $ \tilde{\omega}_{0}=\omega_{0}+\frac{1}{2}N U_{0} $ is the Stark-shifted cavity frequency due to the presence of the BEC, $ \Omega_{c}=4\omega_{R}+\omega_{sw} $ is the frequency of the Bogoliubov mode, $ \zeta=\frac{1}{2}\sqrt{N}U_{0} $ is the optomechanical coupling between the Bogoliubov and the optical modes, and $ \omega_{sw}=8\pi\hbar a_{s}N/m_{a}Lw^2 $ is the \textit{s}-wave scattering frequency of the atomic collisions.

By introducing the Bogoliubov mode quadratures $Q_{c}=(c+c^{\dagger})/\sqrt{2}$ and $P_{c}=(c-c^{\dagger})/\sqrt{2}i$, the Hamiltonian $ H_{c} $ can be written as
\begin{equation}\label{hc}
H_{c}=\frac{1}{2}\hbar\Omega_{c}^{(+)} Q_{c}^{2}+\frac{1}{2}\hbar\Omega_{c}^{(-)} P_{c}^{2}+\hbar\zeta a^{\dagger}a Q_{c},
\end{equation}
where $ \Omega_{c}^{(\pm)}=\Omega_{c}\pm\frac{1}{2}\omega_{sw} $. Now, if we define new atomic quadratures as $ Q=\chi Q_{c} $ and $ P=(1/\chi)P_{c} $ where $ \chi=\big(\frac{\Omega_{c}^{(+)}}{\Omega_{c}^{(-)}}\big)^{1/4} $, the Hamiltonian $ H_{c} $ takes the following form
\begin{equation}\label{Hc}
H_{c}=\frac{1}{2}\hbar\omega_{c}(P^2+Q^2{})+\hbar\xi_{c}a^{\dagger}a Q.
\end{equation}
As is seen from Eq.(\ref{Hc}), the Bogoliubov mode of the BEC behaves as a simple harmonic oscillator with frequency $ \omega_{c}=\sqrt{\Omega_{c}^{(+)}\Omega_{c}^{(-)}} $ which couples to the radiation pressure of the optical field with the optomechanical strength $ \xi_{c}=\frac{1}{\chi}\zeta $.

In the rotating frame with frequency $ \omega_{p}+\dot{\phi} $ the total Hamiltonian of Eq.(\ref{subH}) reads
\begin{eqnarray}\label{H}
H&=&\hbar(\delta_{c}-\dot{\phi})a^{\dagger}a+i\hbar\eta(a-a^{\dagger})+\frac{1}{2}\hbar\omega_{m}(p^{2}+q^{2})\nonumber\\
&&-\hbar\xi_{m} a^{\dagger}a q +\frac{1}{2}\hbar\omega_{c}(P^2+Q^2)+\hbar\xi_{c}a^{\dagger}a Q,
\end{eqnarray}
where $\delta_{c}=-\Delta_{c}+\frac{1}{2}N U_{0}$ is the effective Stark-shifted detuning and $ \Delta_{c}=\omega_{p}-\omega_{0} $.
The Hamiltonian of Eq.(\ref{H}) corresponds to a two-mode optomechanical system with the first mechanical mode $ (p,q) $ and the second atomic (bogoliubov) mode $ (P,Q) $ interacting with the radiation pressure of the cavity. In this form of the Hamiltonian the effect of atom-atom interaction has been coded in both the frequency of the Bogoliubov mode, i.e.,  $ \omega_{c}=\sqrt{(4\omega_{R}+\frac{1}{2}\omega_{sw})(4\omega_{R}+\frac{3}{2}\omega_{sw})} $ and the optomechanical coupling constant $ \xi_{c}=\frac{1}{\chi}\zeta $ through the coefficient $ \chi=(\frac{4\omega_{R}+\frac{3}{2}\omega_{sw}}{4\omega_{R}+\frac{1}{2}\omega_{sw}})^{\frac{1}{4}} $.

\section{The classical model of the LPN}
The laser phase variation rate, i.e., $ \delta\psi=\dot{\phi} $ corresponds to a classical stochastic process which is described by the following equation \cite{Kennedy,Abdi2011}
\begin{equation}\label{psi}
\delta\ddot{\psi}+\tilde{\gamma}\delta\dot{\psi}+\omega_{N}^{2}\delta\psi=\omega_{N}^{2}\sqrt{2\Gamma_{L}}\delta\epsilon(t),
\end{equation}
where $ \omega_{N} $ and $ \tilde{\gamma} $ are, respectively, the central frequency and the bandwidth of the zero-mean noise $ \delta\psi $ while $ \Gamma_{l} $ is the laser linewidth. $ \delta\epsilon(t) $ is a classical white noise with correlation $ \langle\epsilon(t)\epsilon(t^{\prime})\rangle_{cl}=\delta(t-t^{\prime}) $ which is the input noise for the linear stochastic differential equation (\ref{psi}). Based on the theory of classical stochastic processes \cite{papoulis} the power spectrum of the output noise $ \delta\psi(t) $ is determined by the equation $ S_{\delta\psi}(\omega)=|h(\omega)|^{2} S_{\delta\epsilon}(\omega) $, where $ S_{\delta\epsilon}(\omega)=1 $ is the power spectrum of the input white noise and $ h(\omega)=\frac{\omega_{N}^{2}\sqrt{2\Gamma_{L}}}{(\omega^{2}-\omega_{N}^{2})+i\tilde{\gamma}\omega} $ is the linear response function of Eq.(\ref{psi}). Therefore the power spectrum of the output noise $ \delta\psi(t) $ is
\begin{equation}
S_{\delta\psi}(\omega)=\frac{2\Gamma_{l}\omega_{N}^{4}}{(\omega^{2}-\omega_{N}^{2})^{2}+\tilde{\gamma}^{2}\omega^{2}}.
\end{equation}

Now, by defining the auxiliary stochastic variable $ \delta\theta(t)=\frac{1}{\omega_{N}}\delta\dot{\psi} $, the second order differential equation (\ref{psi}) can be written as the following system of first-order differential equations
\begin{subequations}
\begin{eqnarray}
\delta\dot{\psi}&=&\omega_{N}\delta\theta\label{dpsi},\\
\delta\dot{\theta}&=&-\omega_{N}\delta\psi-\tilde{\gamma}\delta\theta+\omega_{N}\sqrt{2\Gamma_{l}}\delta\epsilon\label{dtheta}.
\end{eqnarray}
\end{subequations}

\section{Dyanamics of The System}
For a complete description of the dynamics of the system in the presence of the LPN we should attach the classical noise [Eqs.(\ref{dpsi},\ref{dtheta})] to the linearized Heisenberg-Langevin equations derived from the Hamiltonian of Eq.(\ref{H}). The degrees of freedom of the system can be represented by the vector $u=[q,p,X,Y,Q,P,\delta\psi,\delta\theta]^{T}$ where $X=(a+a^{\dagger})/\sqrt{2}$ and $Y=(a-a^{\dagger})/\sqrt{2}i$ are the optical field quadratures. If the pumping laser is intense enough the quantum operators can be linearized and expanded around their respective classical mean values $u_{s,j}$ as $u_{j}=u_{s,j}+\delta u_{j}$ where $\delta u_{j}$ are zero-mean fluctuation operators ($ j=1,...,6 $). The last two variables ($ \delta u_{7}=\delta\psi $ and $ \delta u_{8}=\delta\theta $) are the classical fluctuations of the laser phase noise and have zero mean values.

In the frame rotating at the fluctuating instantaneous laser frequency $ \omega_{p}+\dot{\phi} $ the phase of the stationary state mean value of the the optical field $ \alpha $ is not random \cite{Abdi2011} while in a frame rotating at the fixed frequency $ \omega_{p} $ it is random \cite{Rabl2011,Phelps}. Since we have chosen the frame rotating at $ \omega_{p}+\dot{\phi} $ the optical mode mean value is given by the standard expressions that are valid in the absence of laser noise, i.e., $ \alpha=-\eta/(\kappa+i\Delta_{d}) $, where $ \Delta_{d} $ is the effective detuning given by $ \Delta_{d}=\delta_{c}-(\frac{\xi_{m}^2}{\omega_{m}}+\frac{\xi_{c}^2}{\omega_{c}})|\alpha|^2 $.

The dynamics of the quantum fluctuations of the system can be described by the linearized QLEs which are coupled to the classical noise equations [Eqs.(\ref{dpsi},\ref{dtheta})]. The linearized QLEs can be written in the compact matrix form
\begin{equation}\label{uA}
\delta\dot{u}(t)=A \delta u(t)+n(t),
\end{equation}
where $\delta u=[\delta q,\delta p,\delta X,\delta Y,\delta Q,\delta P,\delta\psi,\delta\theta]^{T}$ is the vector of continuous variable fluctuations and
\begin{equation}\label{nt}
n(t)=[0,\delta p_{in},\sqrt{2\kappa}\delta X_{in},\sqrt{2\kappa}\delta Y_{in},0,\delta P_{in},0,\omega_{N}\sqrt{2\Gamma_{l}}\epsilon]^{T},
\end{equation}
is the corresponding vector of noises. The optical field input-noise operators are $\delta X_{in}=(\delta a_{in}+\delta a_{in}^{\dagger})/\sqrt{2}$ and $\delta Y_{in}=(\delta a_{in}-\delta a_{in}^{\dagger})/\sqrt{2}i$ where $\delta a_{in}(t)$ satisfies the Markovian correlation functions, i.e., $\langle\delta a_{in}(t)\delta a_{in}^{\dagger}(t^{\prime})\rangle=(n_{ph}+1)\delta(t-t^{\prime})$, $\langle\delta a_{in}^{\dagger}(t^{\prime})\delta a_{in}(t)\rangle=n_{ph}\delta(t-t^{\prime})$ with the average thermal photon number $n_{ph}$ which is nearly zero at optical frequencies \cite{Gardiner}. We also assume that the Brownian noises $ \delta p_{in} $ and $ \delta P_{in} $ affecting, respectively, the mechanical and the Bogoliubov mode of the BEC have Markovian behavior which is valid for oscillators with high quality factors \cite{K Zhang, dalafi2}. Their correlation functions can be written as
\begin{subequations}
\begin{eqnarray}
\langle\delta p_{in}(t)\delta p_{in}(t^{\prime})\rangle&=&\gamma_{m}(2n_{m}+1)\delta(t-t^{\prime}),\\
\langle\delta P_{in}(t)\delta P_{in}(t^{\prime})\rangle&=&\gamma_{c}(2n_{c}+1)\delta(t-t^{\prime}),
\end{eqnarray}
\end{subequations}
where $ \gamma_{m} $ and $ \gamma_{c} $ are the damping rates of the mechanical and the Bogoliubov modes respectively, and $ n_{m}=[\exp(\hbar\omega_{m}/k_{B}T)-1]^{-1} $ and $ n_{c}=[\exp(\hbar\omega_{c}/k_{B}T)-1]^{-1} $ are the thermal excitations of the two modes. The noise sources are assumed uncorrelated for the different modes of the system.

The $8\times8$ drift matrix $A$ is given by
\begin{equation}
\left(\begin{array}{cccccccc}
0 & \omega_{m} & 0 & 0 & 0 & 0 & 0 & 0 \\
-\omega_{m} & -\gamma_{m} & G_{Rm} & G_{Im} & 0 & 0 & 0 & 0 \\
-G_{Im} & 0 & -\kappa & \Delta_{d} & G_{Ic} & 0 & -\sqrt{2}\alpha_{I} & 0 \\
G_{Rm} & 0 & -\Delta_{d} & -\kappa & -G_{Rc} & 0 & \sqrt{2}\alpha_{R} & 0\\
0 & 0 & 0 & 0 & 0 & \omega_{c} & 0 & 0\\
0 & 0 & -G_{Rc} & -G_{Ic} & -\omega_{c} & -\gamma_{c} & 0 & 0\\
0 & 0 & 0 & 0 & 0 & 0 & 0 & \omega_{N}\\
0 & 0 & 0 & 0 & 0 & 0 & -\omega_{N} & -\tilde{\gamma}
  \end{array}\right)
\label{A}
\end{equation}
where the parameters are defined in terms of the real ($ \alpha_{R} $) and imaginary ($ \alpha_{I} $) values of the complex amplitude $ \alpha $ as $ G_{Rm}=\sqrt{2}\alpha_{R}\xi_{m} $, $ G_{Im}=\sqrt{2}\alpha_{I}\xi_{m} $, $ G_{Rc}=\sqrt{2}\alpha_{R}\xi_{c} $ and $ G_{Ic}=\sqrt{2}\alpha_{I}\xi_{c} $. 

The system of linearized dynamical equations given by Eq.(\ref{uA}) with the drift matrix (\ref{A}) corresponds to the following Hamiltonian
\begin{eqnarray}\label{dH}
\delta H&=&-\sqrt{2}\hbar(\alpha_{R}\delta X+\alpha_{I}\delta Y)(\xi_{m}\delta q-\xi_{c}\delta Q+\delta\psi)
\nonumber\\
&&\frac{1}{2}\hbar\Delta_{d}(\delta X^{2}+\delta Y^{2})+\frac{1}{2}\hbar\omega_{m}(\delta q^{2}+\delta p^{2})
\nonumber\\
&&+\frac{1}{2}\hbar\omega_{c}(\delta Q^{2}+\delta P^{2})+\frac{1}{2}\omega_{N}(\delta\psi^{2}+\delta\theta^{2}).
\end{eqnarray}
According to this Hamiltonian, the optical mode $ (\delta X, \delta Y) $ with the effective frequency $ \Delta_{d} $ interacts with both the mechanical mode $ (\delta q, \delta p) $ and the Bogoliubov mode $ (\delta Q, \delta P) $ with respective frequencies $ \omega_{m} $ and $ \omega_{c} $ through the term given in the first line of Eq.(\ref{dH}). The last term in Eq.(\ref{dH}) indicates a classical mode $ (\delta\psi, \delta\theta) $ with frequency $ \omega_{N} $ corresponding to the classical fluctuations of the laser phase. This classical term is responsible for the dynamics of $ \delta\psi $ and $ \delta\theta $ given by Eqs.(\ref{dpsi},\ref{dtheta}).

As is seen from Eq.(\ref{dH}), the classical noise $ \delta\psi $ drives the optical field through the term $ -\sqrt{2}\hbar\delta\psi(\alpha_{R}\delta X+\alpha_{I}\delta Y) $, i.e., it drives the $ \delta X $ quadrature with amplitude $ \alpha_{R} $ and also drives the $ \delta Y $ quadrature with amplitude $ \alpha_{I} $. Therefore, the greater the intracavity photon number the more the effect of LPN on the system.

In the next section we will see how the cavity field behaves as an optical spring which couples the two other modes in the adiabatic approximation.

\section{adiabatic elimination of the optical mode}
If the cavity decay rate $ \kappa $ is much greater than the decay rates of the mechanical and the Bogoliobov modes and the optomechanical couplings, i.e., $ \kappa\gg \gamma_{m}, \gamma_{c}, \xi_{m}, \xi_{c} $, the cavity field may adiabatically be eliminated on time scales greater than $ \kappa^{-1} $. For this purpose, we should find the steady-state solutions of the set of equations $ \delta \dot{X}=0, \delta \dot{Y}=0 $ which leads to a set of algebraic equations. By solving these algebraic equations the quadratures of the optical mode are obtained as follows
\begin{subequations}
\begin{eqnarray}
\delta X&=&\frac{\sqrt{2}(\alpha_{R}\Delta_{d}-\alpha_{I}\kappa)}{\kappa^{2}+\Delta_{d}^{2}}(\xi_{m}\delta q-\xi_{c}\delta Q+\delta\psi)\nonumber\\
&&+\frac{\sqrt{2\kappa}}{\kappa^2+\Delta_{d}^{2}}(\kappa\delta X_{in}+\Delta_{d}\delta Y_{in}),\label{dX}\\
\delta Y&=&\frac{\sqrt{2}(\alpha_{I}\Delta_{d}+\alpha_{R}\kappa)}{\kappa^{2}+\Delta_{d}^{2}}(\xi_{m}\delta q-\xi_{c}\delta Q+\delta\psi)\nonumber\\
&&-\frac{\sqrt{2\kappa}}{\kappa^2+\Delta_{d}^{2}}(\Delta_{d}\delta X_{in}-\kappa\delta Y_{in}).\label{dY}
\end{eqnarray}
\end{subequations}

Substituting Eqs.(\ref{dX}) and (\ref{dY}) into the other equations of the system, Eq.(\ref{uA}), the following new set of Heisenberg-Langevin equations will be obtained
\begin{eqnarray}\label{ad HL}
\delta\dot{q}&=&\omega_{m}\delta p,\nonumber\\
\delta\dot{p}&=&-(\omega_{m}-\nu_{m})\delta q-\gamma_{m}\delta p-\sqrt{\nu_{m}\nu_{c}}\delta Q+\frac{\nu_{m}}{\xi_{m}}\delta\psi+\delta n_{2}(t),\nonumber\\
\delta\dot{Q}&=&\omega_{c}\delta P,\nonumber\\
\delta\dot{P}&=&-\sqrt{\nu_{m}\nu_{c}}\delta q-(\omega_{c}-\nu_{c})\delta Q-\gamma_{c}\delta P+\frac{\nu_{c}}{\xi_{c}}\delta\psi+\delta n_{4}(t),\nonumber\\
\end{eqnarray}
which are again coupled to the classical equations of laser phase fluctuations, i.e., Eqs.(\ref{dpsi}) and (\ref{dtheta}). Here, $ \nu_{m}=g_{m}^{2}\frac{\Delta_{d}}{\kappa^{2}+\Delta_{d}^{2}} $ and $ \nu_{c}=g_{c}^{2}\frac{\Delta_{d}}{\kappa^{2}+\Delta_{d}^{2}} $ are, respectively, the frequency shifts of the mechanical and the Bogoliubov modes due to the radiation pressure where
 $ g_{m}=\sqrt{2}\xi_{m}|\alpha| $ and $ g_{c}=\sqrt{2}\xi_{c}|\alpha| $. Besides, the quantum noises $ \delta n_{2}(t)=\delta p_{in}+\xi_{m}\delta Z(t) $ and $ \delta n_{4}(t)=\delta P_{in}-\xi_{c}\delta Z(t) $ where
\begin{equation}\label{dZ}
\delta Z=\frac{2\sqrt{\kappa}}{\kappa^2+\Delta_{d}^{2}}[\kappa(\alpha_{R}\delta X_{in}+\alpha_{I}\delta Y_{in})+\Delta_{d}(\alpha_{R}\delta Y_{in}-\alpha_{I})\delta X_{in}],
\end{equation}
is the quantum noise due to the optical field that affects both the mechanical and the Bogoliubov modes.

It can be shown that the set of dynamical Eqs.(\ref{ad HL}), (\ref{dpsi}) and (\ref{dtheta}) are governed by the following effective Hamiltonian
\begin{eqnarray}\label{dhad}
\delta H_{ad}&=&\frac{1}{2}\hbar[(\omega_{m}-\nu_{m})\delta q^{2}+\omega_{m}\delta p^{2})]+\hbar\sqrt{\nu_{m}\nu_{c}}\delta q\delta Q\nonumber\\
&&\frac{1}{2}\hbar[(\omega_{c}-\nu_{c})\delta Q^{2}+\omega_{c}\delta P^{2})]+\hbar\delta\psi(\frac{\nu_{c}}{\xi_{c}}\delta Q-\frac{\nu_{m}}{\xi_{m}}\delta q)\nonumber\\
&&+\frac{1}{2}\omega_{N}(\delta\psi^{2}+\delta\theta^{2}).
\end{eqnarray}
Using the transformations $ \delta\tilde{q}=\tilde{\chi}_{m}\delta q $, $ \delta\tilde{p}=1/\tilde{\chi}_{m}\delta p $, $ \delta\tilde{Q}=\tilde{\chi}_{c}\delta Q $ and $ \delta\tilde{P}=1/\tilde{\chi}_{c}\delta P $ where $ \tilde{\chi}_{i}=(\frac{\omega_{i}-\nu_{i}}{\omega_{i}})^{1/4} $ for $ i=m, c $, the Hamiltonian of Eq.(\ref{dhad}) can be written as
\begin{eqnarray}\label{dHad}
\delta H_{ad}&=&\frac{1}{2}\hbar\tilde{\omega}_{m}(\delta\tilde{q}^{2}+\delta\tilde{p}^{2})+\frac{1}{2}\hbar\tilde{\omega}_{c}(\delta\tilde{Q}^{2}+\delta\tilde{P}^{2})+\hbar\tilde{G}_{mc}\delta\tilde{q}\delta\tilde{Q}\nonumber\\
&&+\hbar\delta\psi(\tilde{r}_{c}\delta\tilde{Q}-\tilde{r}_{m}\delta\tilde{q})+\frac{1}{2}\omega_{N}(\delta\psi^{2}+\delta\theta^{2}),
\end{eqnarray}
where $ \tilde{\omega}_{m}=\sqrt{\omega_{m}(\omega_{m}-\nu_{m})} $ and $ \tilde{\omega}_{c}=\sqrt{\omega_{c}(\omega_{c}-\nu_{c})} $ are, respectively, the effective frequency of the Bogoliubov and the mechanical modes. Furthermore, the two modes are coupled to each other through the term $ \delta\tilde{q}\delta\tilde{Q} $ with the effective coupling parameter
\begin{equation}
\tilde{G}_{mc}=\frac{\sqrt{\nu_{m}\nu_{c}}}{\tilde{\chi}_{m}\tilde{\chi}_{c}}=\sqrt{\nu_{m}\nu_{c}}\big[\frac{\omega_{m}\omega_{c}}{(\omega_{m}-\nu_{m})(\omega_{c}-\nu_{c})}\big]^{1/4}.
\end{equation}
Based on these results, the optical mode behaves as a spring which couples the amplitude quadratures of the mechanical and the Bogoliubov modes through the effective coupling parameter $ \tilde{G}_{mc} $. Besides, the quantum noise of the optical mode appears as the effective noise $ \delta Z(t) $ given by Eq.(\ref{dZ}) which affects the phase quadratures of the two modes. 

The fourth term in Eq.(\ref{dHad}) indicates the effect of the classical laser phase fluctuations on the two quantum modes. As is seen, the classical phase noise drives the amplitude quadratures of the mechanical and the Bogoliubov modes with rates $ \tilde{r}_{m}\delta\psi $ and $ \tilde{r}_{c}\delta\psi $, respectively where the dimensionless phase noise coupling parameters $ \tilde{r}_{m,c} $ have been defined as
\begin{subequations}
\begin{eqnarray}
\tilde{r}_{m}&=&\frac{1}{\tilde{\chi}_{m}}\frac{\nu_{m}}{\xi_{m}},\label{rm}\\
\tilde{r}_{c}&=&\frac{1}{\tilde{\chi}_{c}}\frac{\nu_{c}}{\xi_{c}}.\label{rc}
\end{eqnarray}
\end{subequations}

Now, we would like to examine the behavior of the effective frequencies and the coupling parameter in terms of the effective detuning $ \Delta_{d} $. We analyze our results based on the experimentally feasible parameters given in \cite{Ritter Appl. Phys. B, Brenn Science},i.e., we assume there are $ N=10^5 $ Rb atoms inside an optical cavity of length $ L=187 \mu$m with bare frequency $ \omega_{c}=2.41494\times 10^{15} $Hz corresponding to a wavelength of $ \lambda=780 $nm and damping rate $ \kappa=2\pi\times 1.3 $MHz. The cavity mode is coherently driven at amplitude $ \eta $ by a pump laser with frequency $ \omega_{p} $ through the fixed end mirror. The atomic $ D_{2} $ transition corresponding to the atomic transition frequency $ \omega_{a}=2.41419\times 10^{15} $Hz couples to the mentioned mode of the cavity. The atom-field coupling strength $ g_{0}=2\pi\times 14.1 $MHz and the recoil frequency of the atoms is $ \omega_{R}=23.7 $KHz. The movable end mirror with mass $ m=10^{-9}  $g and damping rate $ \gamma_{m}=2\pi\times 100 $Hz oscillates with frequency $ \omega_{m}=10^5 $Hz. 

The Bogoliubov mode of the BEC oscillates with frequency $ \omega_{c}=\sqrt{(4\omega_{R}+\frac{1}{2}\omega_{sw})(4\omega_{R}+\frac{3}{2}\omega_{sw})} $ and damps with rate $ \gamma_{c}=0.001\kappa $. Therefore, $ \omega_{c} $ increases monotonically in terms of $ \omega_{sw} $. Since the \textit{s}-wave scattering frequency $ \omega_{sw} $ is controllable through the transverse trapping frequency $\omega_{\mathrm{\perp}}$, the frequency $ \omega_{c} $ can be manipulated experimentally. By fixing $ \omega_{sw}=0.2\omega_{R} $ one can adjust the frequency of the Bogoliubov mode very near to that of the mechanical mode, i.e.,  $ \omega_{c}\approx\omega_{m} $. For this set of parameters the system is in the Doppler regime where $ \kappa\gg\omega_{c}, \omega_{m} $.

\begin{figure}[ht]
\centering
\includegraphics[width=2.7in]{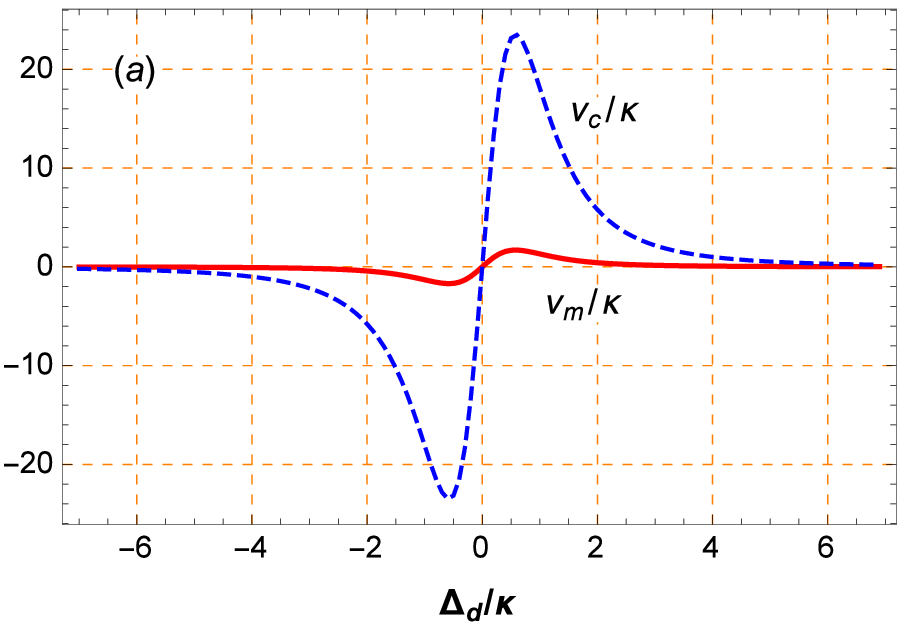}
\includegraphics[width=2.7in]{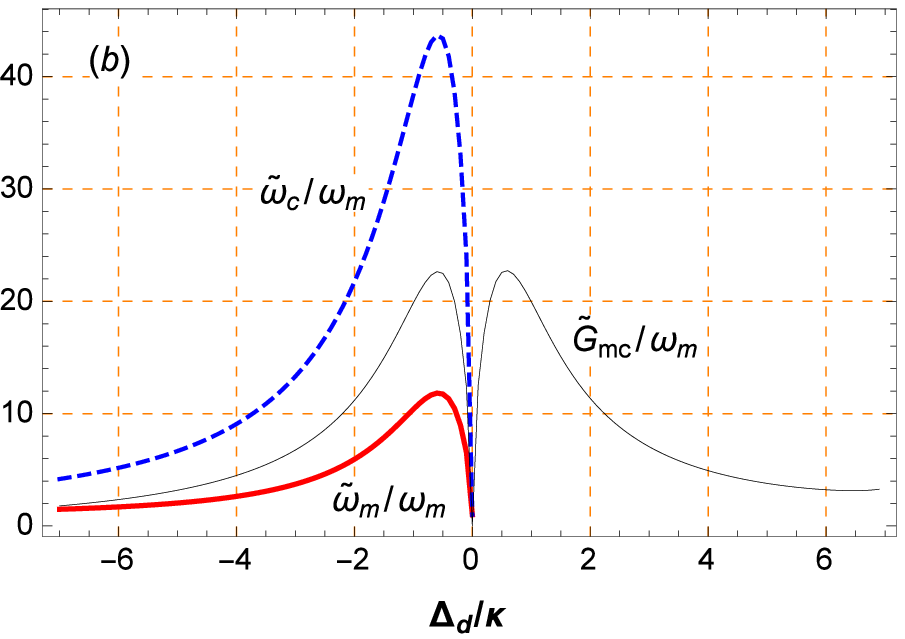}
\caption{
(Color online) (a) The normalized frequency shifts of the mechanical ($ \nu_{m}/\kappa $, red solid line) and the Bogoliubov ($ \nu_{c}/\kappa $, blue dashed line) modes versus the normalized effective detuning $ \Delta_{d}/\kappa $, and (b)the normalized effective frequencies of the mechanical ($ \tilde{\omega}_{m}/\omega_{m} $, red solid line) and the Bogoliubov ($ \tilde{\omega}_{c}/\omega_{m} $, blue dashed line) modes and the normalized effective coupling parameter $ \tilde{G}_{mc}/\omega_{m} $ (black thin line) versus normalized effective detuning $ \Delta_{d}/\kappa $. The parameters are $ \eta=30\kappa $, $ \omega_{c}\approx\omega_{m}=10^{5} $Hz, $ \chi_{c}=0.2\kappa $, and $ \chi_{m}=0.05\kappa $.}
\label{fig:fig2}
\end{figure}

\begin{figure}[ht]
\centering
\includegraphics[width=2.7in]{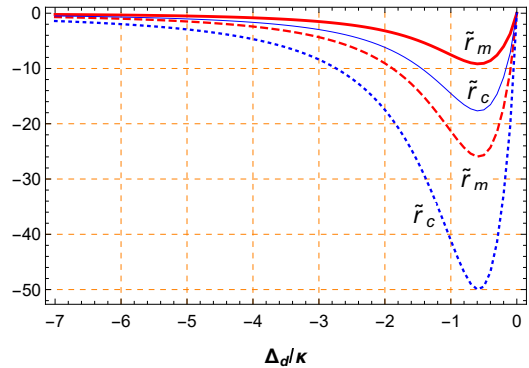} 
\caption{(Color online) The dimensionless phase noise coupling parameters $ \tilde{r}_{m}(\eta=30\kappa) $ (red solid line) , $ \tilde{r}_{m}(\eta=60\kappa) $ (red dashed line), $ \tilde{r}_{c}(\eta=30\kappa) $ (blue thin line) and $ \tilde{r}_{c}(\eta=60\kappa) $ (blue dotted line) versus the normalized effective detuning $ \Delta_{d}/\kappa $ . The parameters are the same as those of Fig.\ref{fig:fig2}.}
\label{fig:fig3}
\end{figure}

In Fig.\ref{fig:fig2} (a) the normalized frequency shifts $ \nu_{c}/\kappa $ (blue dashed line) and $ \nu_{m}/\kappa $ (red solid line) have been plotted versus the normalized effective detuning $ \Delta_{d}/\kappa $. Since the widths of the two curves are the same they never cross each other for $ |\Delta_{d}|<\kappa $. It means that there is no resonance condition in this region. However, in the dispersive regime where $ |\Delta_{d}|\gg\kappa $ the two frequency shifts get near to zero and the resonance condition is fulfilled. In order to see this situation more clearly, in Fig.\ref{fig:fig2} (b) we have plotted the normalized effective frequencies $ \tilde{\omega}_{c}/\omega_{m} $ (blue dashed line) and $ \tilde{\omega}_{m}/\omega_{m} $ (red solid line) versus $ \Delta_{d}/\kappa $. As is seen, the two curves never intersect each other except for very large negative values of $ \Delta_{d} $ where the two effective frequencies coincide and therefore the two oscillators are brought into resonance with each other. Besides, the effective frequencies are undefined in a wide range of $ \Delta_{d}>0 $ where $ \nu_{i}>\omega_{i} $ $ (i=m,c) $ and the effective frequencies $ \tilde{\omega_{i}}=\sqrt{\omega_{i}(\omega_{i}-\nu_{i})} $ are no longer real.

In Fig.\ref{fig:fig2} (b) we have also plotted the normalized effective coupling parameter $ \tilde{G}_{mc}/\omega_{m} $ (black thin line) versus $ \Delta_{d}/\kappa $. The maxima of the curve occur in the regions $ |\Delta_{d}|<\kappa $ where there are no resonances. For larger values of $ |\Delta_{d}| $ the coupling parameter reduces but is still considerable enough to cause the two oscillator to interact with each other effectively even in the dispersive regime.

Finally, in order to see the effect of the LPN on the optomechanical system, in Fig.\ref{fig:fig3} we have plotted the dimensionless phase noise coupling parameters $ \tilde{r}_{c} $ and $ \tilde{r}_{m} $ [Eqs.(\ref{rm},\ref{rc})] versus the normalized detuning $ \Delta_{d}/\kappa $ for two different values of the pump laser rate $ \eta=30\kappa $ (solid lines) and $ \eta=60\kappa $ (dashed and dotted lines).  As is seen, by increasing the pump laser power (or equivalently $ \eta $) the classical noise is injected into the system at faster rates for all values of $ \Delta_{d} $.  Besides, the maxima of $ |\tilde{r}_{m}| $ and $ |\tilde{r}_{m}| $ occur around $ \Delta_{d}\approx-\kappa/2 $ where the oscillators are not in resonance while in the dispersive regime $ (|\Delta_{d}|\gg\kappa) $ both the classical coupling constants get very low (but still nonzero). Therefore, in the dispersive regime where the oscillators are at resonance, the LPN has the least disturbance on the two system modes.

Since the frequency shifts $ \nu_{i} $ are proportional to the intracavity photon number, increasing the intensity of the external pump laser causes the frequency shifts to be increased which in turn leads to the increase of the effective coupling parameter $ \tilde{G}_{mc} $ and also the increase of the dimensionless phase noise coupling parameters $ \tilde{r}_{i} $. Therefore, the more the pump laser intensity the more laser phase noise injected into the system because of the increase of the phase noise coupling parameters $ \tilde{r}_{i} $. In the next section we will see how the interaction of the two oscillators in the dispersive regime can lead to the production of a strong stationary entanglement between them and also we will investigate the effect of LPN as well as atom-atom interaction on the generated bipartite entanglements.

\section{Effects of LPN and atomic collisions on bipartite Entanglements}
In this section we investigate how the LPN and the atom-atom interaction affect the bipartite entanglements of the system. For this purpose, we should find the correlation functions of the system in the stationary state in the regime where the system is stable.

The system is stable only if the real part of all the eigenvalues of the matrix $A$ are negative. These stability conditions can be obtained by using the Routh-Hurwitz criterion \cite{RH}. Due to the linearized dynamics of the fluctuations and since all noises are Gaussian the steady state is a zero-mean Gaussian state which is fully characterized by the $8\times8$ stationary correlation matrix (CM) $V$, with components $V_{ij}=\langle \delta u_i(\infty)\delta u_j(\infty)+\delta u_j(\infty)\delta u_i(\infty)\rangle/2 $. Using the QLEs, one can show that $ V $ fulfills the  Lyapunov equation \cite{Genes2008}
\begin{equation}\label{lyap}
AV+VA^T=-D,
\end{equation}
where
\begin{equation}\label{D}
 D=\mathrm{Diag}[0,\gamma_{m}^{\prime},\kappa,\kappa,0,\gamma_{c}^{\prime},0,2\Gamma_{l}\omega_{N}^2],
\end{equation}
is the diffusion matrix with $ \gamma_{m}^{\prime}=\gamma_{m}(2n_{m}+1) $ and $ \gamma_{c}^{\prime}=\gamma_{c}(2n_{c}+1) $. Eq.(\ref{lyap}) is linear in $V$ and can straightforwardly be solved. However, the explicit form of $ V $ is complicated and is not reported here.

\begin{figure}[ht]
\centering
\includegraphics[width=2.5in]{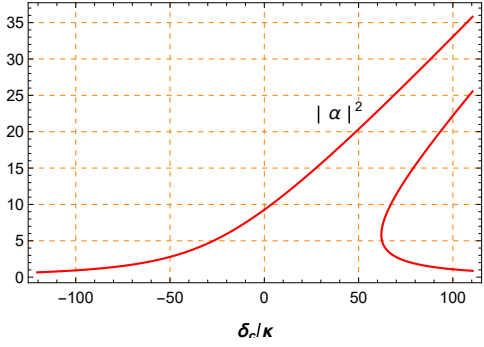} 
\caption{(Color online) Mean intracavity photon number $\alpha^2$ versus the normalized cavity-pump detuning $\delta_{c}/\kappa$, for pumping rate $ \eta=100\kappa $ where $ \kappa=2\pi\times 1.3 $MHz is the damping rate of the cavity. The values of other parameters are $ L=187 \mu$m, $ \lambda=780 $nm, $ m=10^{-9} $g, $ \gamma_{m}=2\pi\times 100 $Hz, and $ \omega_{m}=10^5 $Hz. The cavity contains $ N=10^5 $ Rb atoms with $ \omega_{sw}=0.2\omega_{R} $ and $ \gamma_{c}=0.001\kappa $. }
\label{fig:fig4}
\end{figure}

By obtaining the intracavity amplitude $ \alpha $ the drift matrix elements are determined. Now, by solving the Lyapunov equation [Eq.(\ref{lyap})] we can obtain the correlation matrix $V$ which gives us the second-order correlations of the fluctuations. The bipartite entanglement can also be calculated by using the logarithmic negativity \cite{eis}:
\begin{equation}\label{en}
E_N=\mathrm{max}[0,-\mathrm{ln} 2 \eta^-],
\end{equation}
where  $\eta^{-}\equiv2^{-1/2}\left[\Sigma(\mathcal{V}_{bp})-\sqrt{\Sigma(\mathcal{V}_{bp)}^2-4 \mathrm{det} \mathcal{V}_{bp}}\right]^{1/2}$  is the lowest symplectic eigenvalue of the partial transpose of the $4\times4$ CM, $\mathcal{V}_{bp}$, associated with the selected bipartition, obtained by neglecting the rows and columns of the uninteresting modes
\begin{equation}\label{bp}
\mathcal{V}_{bp}=\left(
     \begin{array}{cc}
     \mathcal{B}&\mathcal{C}\\
      \mathcal{C}^{T}&\mathcal{B}^{\prime}\\
       \end{array}
   \right),
\end{equation}
and $\Sigma(\mathcal{V}_{bp})=\mathrm{det} \mathcal{B}+\mathrm{det} \mathcal{B}^{\prime}-2\mathrm{det} \mathcal{C}$.

In Fig.\ref{fig:fig4}, we have plotted the mean intracavity photon number versus the normalized cavity-pump detuning $\delta_{c}/\kappa$ when the cavity is pumped by a laser at rate $ \eta=100\kappa $. For the chosen set of parameters when $ -150\leq\delta_{c}/\kappa\leq 150 $, the effective frequency of the cavity mode is in the range $ -14000\leq\Delta_{d}/\omega_{m}\leq -2000 $. It means that the system is in the dispersive regime where $ |\Delta_{d}|\gg\omega_{m}, \kappa $. Besides, the cavity damping rate is much greater than other damping rates as well as the optomechanical couplings, i.e., $ \kappa\gg\gamma_{m}, \gamma_{c}, \xi_{b}, \xi_{c} $. 

\begin{figure}[ht]
\centering
\includegraphics[width=2.5in]{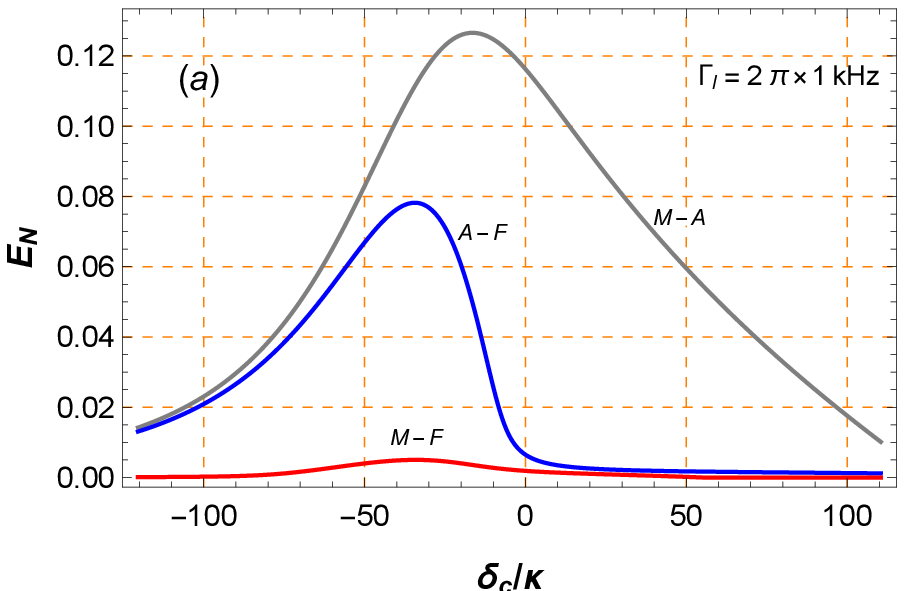}
\includegraphics[width=2.5in]{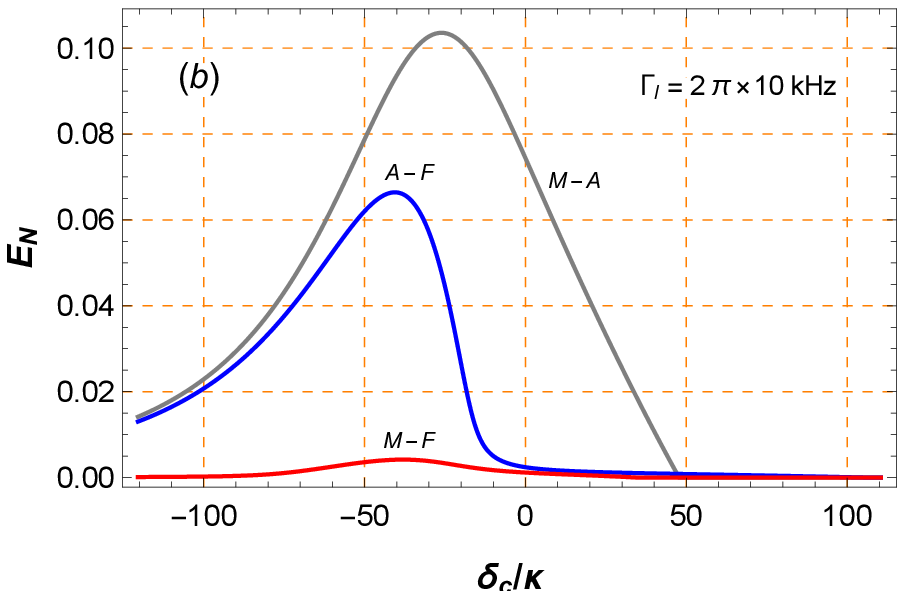}
\includegraphics[width=2.5in]{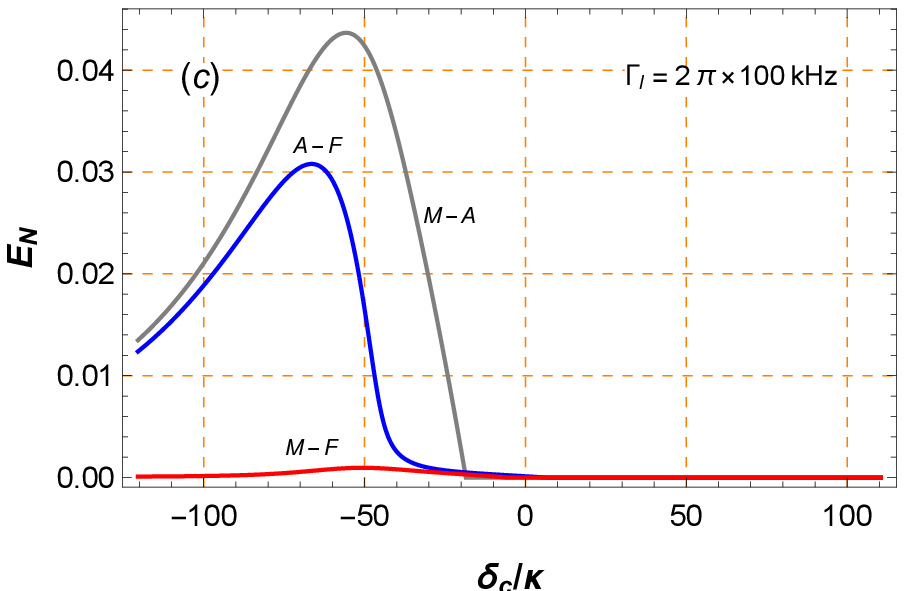}
\caption{
(Color online) The bipartite mirror-atom (M-A, gray line), atom-field (A-F, blue line) and mirror-field (M-F, red line) entanglements versus the normalized detuning $\delta_{c}/\kappa$ for three different values of the laser linewidth: (a) $ \Gamma_{l}=2\pi\times 1 $kHz, (b) $ \Gamma_{l}=2\pi\times 10 $kHz, and (c) $ \Gamma_{l}=2\pi\times 100 $kHz. The laser noise parameters are $ \omega_{N}=2\pi\times 140 $kHz, $ \tilde{\gamma}=\omega_{N}/2 $. The other parameters are the same as those of Fig.\ref{fig:fig4}.  }
\label{fig:fig5}
\end{figure}

In Fig.\ref{fig:fig5} we have plotted the bipartite entanglements between the quantun modes of the system for three different values of the laser linewidth when the cavity is pumped at rate $ \eta=100\kappa $ (just like Fig.\ref{fig:fig4}). Here, mirror-atom (M-A, gray line), atom-field (A-F, blue line) and mirror-field (M-F, red line) entanglements have been plotted versus the normalized cavity-pump detuning $\delta_{c}/\kappa$ for three different values of the laser linewidth $ \Gamma_{l}=2\pi\times 1 $kHz [Fig.\ref{fig:fig5}(a)], $ \Gamma_{l}=2\pi\times 10 $kHz [Fig.\ref{fig:fig5}(b)], and $ \Gamma_{l}=2\pi\times 100 $kHz [Fig.\ref{fig:fig5}(c)]. The central frequency and the bandwidth of the zero-mean noise $ \delta\psi $ have been fixed at $ \omega_{N}=2\pi\times 140 $kHz and $ \tilde{\gamma}=\omega_{N}/2 $, respectively. Besides, we have assumed that the temperature of the system is fixed at $ T=0.1\mu $K.

What is more important for us here, is the entanglement produced between the atomic field and the vibrating motion of the mirror, i.e., M-A entanglement. As is seen, at such low temperatures where all thermal noises of the system are nearly negligible , a strong stationary entanglement between the Bogoliubov mode of the BEC and the mirror vibrations can be established when the system is in the dispersive regime. Under these conditions, the only noise that still affects the system is the classical fluctuation of the laser phase which is technically inevitable.

\begin{figure}[ht]
\centering
\includegraphics[width=2.8in]{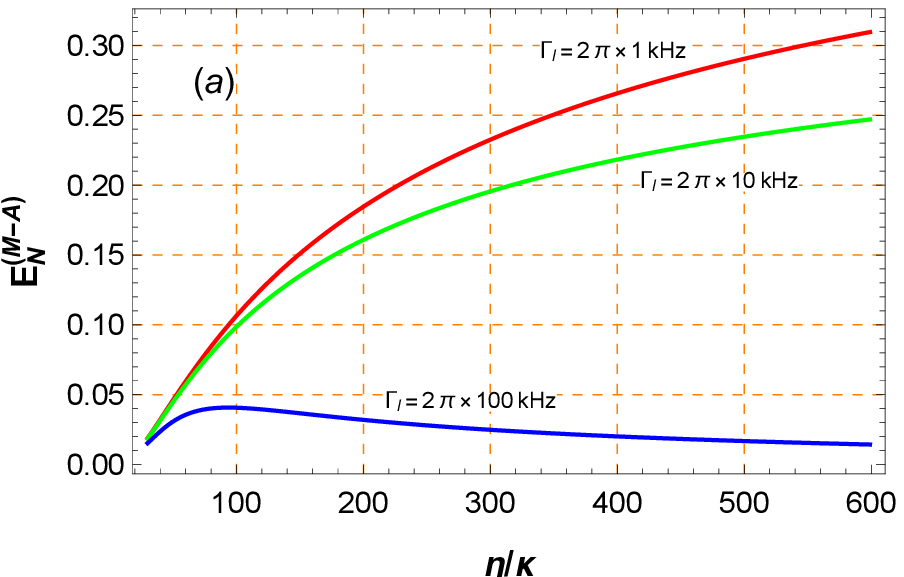} 
\includegraphics[width=2.8in]{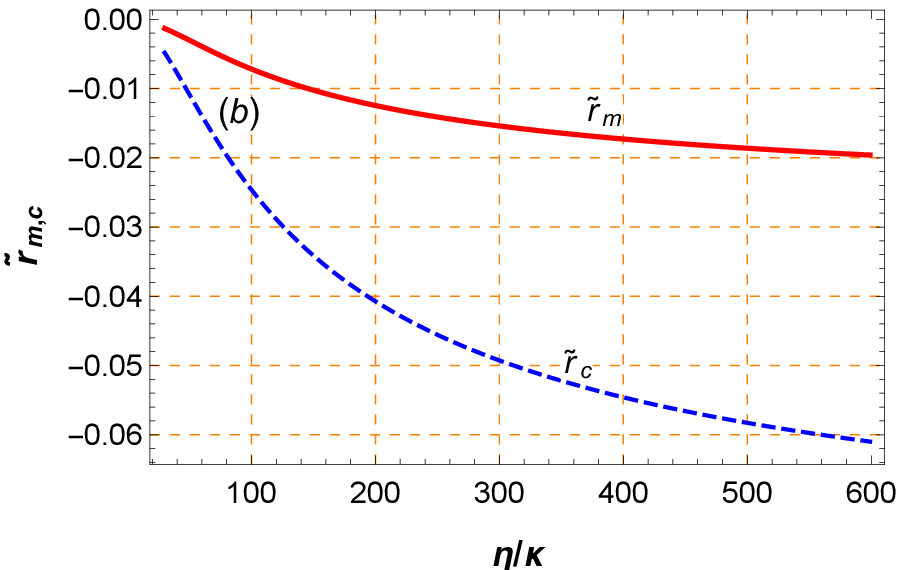}
\caption{(Color online)  (a) The mirror-atom (M-A) entanglement versus the normalized pump laser rate $ \eta/\kappa $  at the cavity-pump detuning $ \delta_{c}=-40\kappa $ for three different values of the laser linewidth $ \Gamma_{l}=2\pi\times 1 $kHz (red line), $ \Gamma_{l}=2\pi\times 10 $kHz (green line) , and $ \Gamma_{l}=2\pi\times 100 $kHz (blue line).  (b) The dimensionless phase noise coupling parameters $ \tilde{r}_{m} $ and $ \tilde{r}_{c} $  versus the normalized pump laser rate $ \eta/\kappa $. The other parameters are the same as those of Fig.\ref{fig:fig4} and Fig.\ref{fig:fig5}.  }
\label{fig:fig6}
\end{figure}

As is seen from Fig.\ref{fig:fig5}(a), for a laser with $ \Gamma_{l}=2\pi\times 1 $kHz there is a strong M-A entanglement for a wide range of $ \delta_{c} $ with a maximum of $E_{N}\approx 0.13 $ at $ \delta_{c}\approx -15\kappa $. The results obtained in Fig.\ref{fig:fig5}(a) are very similar to those in the absence of LPN, i.e., when $ \Gamma_{l}=0 $ (which we have not shown its plot because it is very similar to that of Fig.\ref{fig:fig5}(a)). Therefore, for pump lasers with linewidths $ \Gamma_{l}\leq2\pi\times 1 $kHz the laser phase noise is so low that cannot affect the stationary entanglements of the system. However, for pump lasers with larger linewidth the effect of LPN  is no longer negligible. In Figs.\ref{fig:fig5} (b) and (c) the values of the laser linewidth have been fixed at $ \Gamma_{l}=2\pi\times 10 $kHz and $ \Gamma_{l}=2\pi\times 100 $kHz, respectively. As is seen, for these larger values of the laser linewidth, not only the the maxima of all the bipartite entanglements reduce but also the range of the detuning in which the entanglements have nonzero values gets much narrower.

\begin{figure}[ht]
\centering
\includegraphics[width=2.8in]{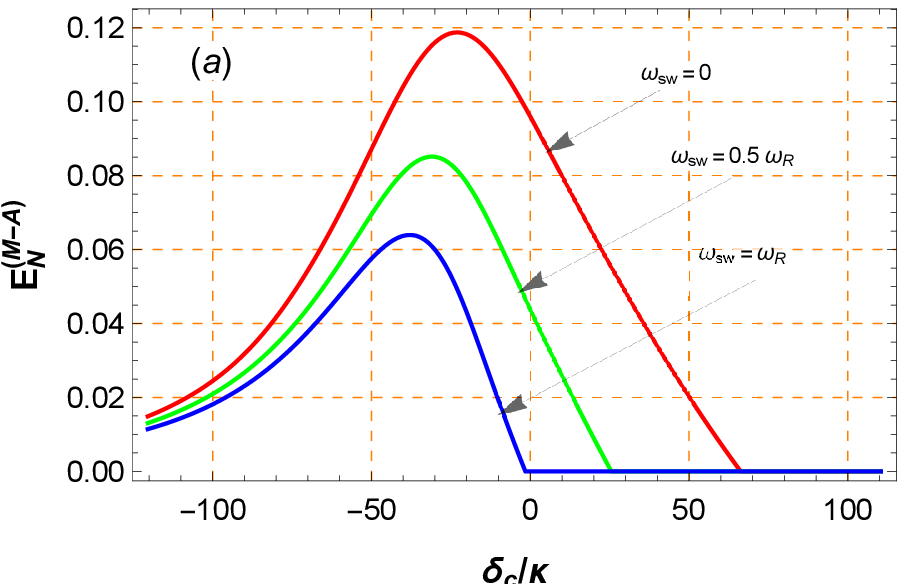}
\includegraphics[width=2.8in]{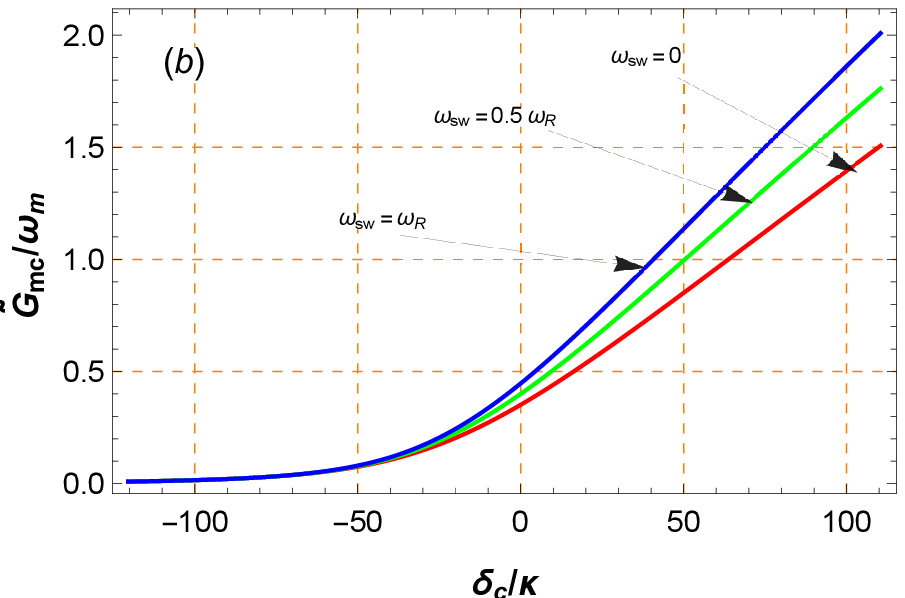}
\includegraphics[width=2.8in]{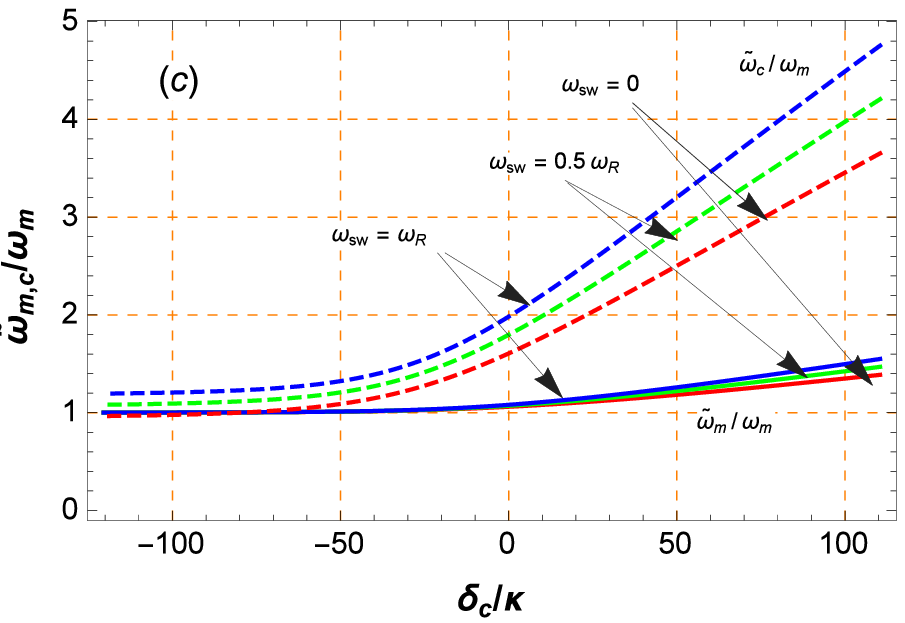}
\caption{
(a) The mirror-atom (M-A) entanglement, (b) the normalized effective coupling parameter $ \tilde{G}_{mc}/\omega_{m} $, and (c) the normalized effective frequencies of the mechanical ($ \tilde{\omega}_{m}/\omega_{m} $, solid lines) and the Bogoliubov ($ \tilde{\omega}_{c}/\omega_{m} $, dashed lines) modes versus the normalized detuninng $ \delta_{c}/\kappa $ for three different values of $ \omega_{sw}=0 $ (red line), $ \omega_{sw}=0.5\omega_{R} $ (green line) and $ \omega_{sw}=\omega_{R} $ (blue line). The laser linewidth is $ \Gamma_{l}=2\pi\times 10 $kHz and the other parameters are the same as those of Fig.\ref{fig:fig4} and Fig.\ref{fig:fig5}.}
\label{fig:fig7}
\end{figure}

In Fig.\ref{fig:fig6}(a) the stationary mirror-atom entanglement has been plotted against the normalized pump laser rate $ \eta/\kappa $  at the cavity-pump detuning $ \delta_{c}=-40\kappa $ for three different values of the laser linewidth $ \Gamma_{l}=2\pi\times 1 $kHz (red line), $ \Gamma_{l}=2\pi\times 10 $kHz (green line) , and $ \Gamma_{l}=2\pi\times 100 $kHz (blue line).  As is seen, for each value of the pump rate $ \eta $ the entanglement is decreased by increasing the laser linewidth. Besides, for lower values of the laser linewidth ($ \Gamma_{l}=2\pi\times 1 $kHz-$ 2\pi\times 10 $kHz)  the M-A entanglement is increased by increasing the pump laser rate (red and green lines) because at such low values of the laser linewidth the effect of phase fluctuations are not very strong. However, for $ \Gamma_{l}=2\pi\times 100 $kHz the M-A entanglement is a little bit increased for laser pump rates $ \eta<80\kappa $ while is decreased for $ \eta>80\kappa $.  

In Fig.\ref{fig:fig6}(b) the dimensionless phase noise coupling parameters $ \tilde{r}_{c} $ (blue dashed line) and $ \tilde{r}_{m} $ (red solid line) [Eqs.(\ref{rm},\ref{rc})] have been plotted versus  the normalized pump laser rate $ \eta/\kappa $. As is seen, the absolute values of them are increased by increasing $ \eta $. Since these phase noise coupling parameters are independent of the laser linewidth ($ \Gamma_{l} $), the phase noise is injected into the system by the same rate for lasers having different linewidths. However,  since the quantum correlations are affected by the laser linewidth, the entanglement is reduced for larger values of the laser linewidth. In other words, by increasing the input power of the pump laser, the noise is injected into the system with faster rate. However, this noise cannot affect the stationary M-A entanglement as far as the laser linewidth is very low. Instead, for higher values of the laser linewidth, the destructive effect of the phase noise dominates and makes the entanglement decrease by increasing the power of the pump laser.

In order to see the effect of the atom-atom interaction on the M-A entanglement, in Fig.\ref{fig:fig7}(a) we have plotted the M-A entanglement versus the normalized cavity-pump detuning $\delta_{c}/\kappa$ for three different values of the \textit{s}-wave scattering frequencies $ \omega_{sw}=0 $ (red line), $ \omega_{sw}=0.5\omega_{R} $ (green line) and $ \omega_{sw}=\omega_{R} $ (blue line). As is seen, the M-A entanglement is reduced by increasing the \textit{s}-wave scattering frequency. To understand why this phenomenon is happening, we have plotted the normalized effective coupling parameter $ \tilde{G}_{mc}/\omega_{m} $ [Fig.\ref{fig:fig7} (b)] and the normalized effective frequencies of the mechanical mode ($ \tilde{\omega}_{m}/\omega_{m} $, solid lines) and of the Bogoliubov mode ($ \tilde{\omega}_{c}/\omega_{m} $, dashed lines) versus the normalized detuninng $ \delta_{c}/\kappa $ [Fig.\ref{fig:fig7} (c)].

Based on these numerical results, increasing the \textit{s}-wave scattering frequency from one hand causes the effective coupling parameter $ \tilde{G}_{mc} $ to be increased a little bit [Fig.\ref{fig:fig7}(b)] and from the other hand, causes the effective frequencies of the two modes ($ \tilde{\omega}_{m} $ and $ \tilde{\omega}_{c} $) to get far away from each other [Fig.\ref{fig:fig7}(c)]. As is seen, the M-A entanglement occurs in the region where the two oscillators are nearly at resonance (where $ \tilde{\omega}_{c}\longrightarrow\tilde{\omega}_{m} $ ) as far as the effective coupling parameter is nonzero. For larger negative values of $ \delta_{c} $ the two modes will be at resonance but the entanglement vanished since the effective coupling parameter tends to zero. 

Therefore, the atom-atom interaction has a dual effect:  firstly, it strengthens the effective coupling parameter between the two modes that should lead to an increase of the entanglement and secondly, it makes the two modes get out of resonance. Since the latter effect dominates the former, the ultimate effect of the atom-atom interaction appears as a reduction of the entanglement. This effect is more considerable for $ \delta_{c}>0 $. Besides, the increase in the effective coupling parameter due to the atom-atom interaction is more considerable for larger values of $ \delta_{c} $ where the system gets near to the bisatbility region and the entanglement tends to zero. 

\section{Conclusions}
In conclusion, we have studied a driven optomechanical cavity with a moving end mirror containing a cigar shaped BEC.  Since in any realistic setup the external laser that pumps the cavity has a nonzero linewidth because of its phase fluctuations, we have taken into account the effect of this classical phase noise in our model.  In the weakly interacting regime, where just the first two symmetric momentum side modes are excited by fluctuations resulting from the atom-light interaction, the BEC can be considered as a one mode oscillator in the Bogoliubov approximation. In this way, the Bogoliubov mode of the BEC is coupled to the optical field through a radiation pressure term which behaves as a secondary mechanical mode relative to the vibrational mode of the mirror.

We have investigated the behavior of the system in the dispersive regime where the effective frequency of the optical mode is much greater than those of the mechanical and the Bogoliubov modes.  Besides, we consider the situation where the damping rate of the cavity is faster than those of the other modes so that the optical field follows the dynamics of the mechanical and atomic oscillators adiabatically. In this way, an effective two-mode Hamiltonian is derived in the adiabatic approximation where the Bogoliubov and mechanical modes are coupled to each other through the mediation of the optical field by an effective coupling parameter. 

It was shown that the LPN in the effective two-mode model appears as a classical stochastic pump term which drives the amplitude quadratures of both the mechanical and the Bogoliubov modes. We have also shown that a strong stationary M-A entanglement can be established just in the dispersive and Doppler regimes where the two modes come into resonance with each other and the effect of laser phase noise gets very low. Meanwhile, we have shown that although the nonlinear effect of atom-atom interaction increases the effective coupling parameter between the two modes, it leads to a reduction in the M-A entanglement because it makes the two modes get out of resonance.

\section*{Acknowledgement}
A.D wishes to thank the Laser and Plasma Research Institute of Shahid Beheshti University for its support.

\bibliographystyle{apsrev4-1}

\end{document}